\documentclass[a4paper,11pt]{article}
\pdfoutput=1 
\usepackage{jcappub} 

\usepackage[T1]{fontenc} 
\usepackage[applemac]{inputenc}

\usepackage{amsmath}
\usepackage{amssymb}
\usepackage{subfig}

 \title{\boldmath Microscopic Picture of Non-Relativistic Classicalons}
 
 \author[a]{Felix Berkhahn,}
  \author[a,b]{Sophia Müller,}
  \author[a,c]{Florian Niedermann}
    \author[a,c]{and  Robert Schneider}
 \affiliation[a]{Arnold Sommerfeld Center for Theoretical Physics, Ludwig-Maximilians-Universit\"at, Theresienstra{\ss}e 37, 80333 Munich, Germany}
\affiliation[b]{Max-Planck-Institute for Physics, Foehringer Ring 6, 80805 Munich, Germany}
  \affiliation[c]{Excellence Cluster Universe, Boltzmannstra{\ss}e 2, 85748 Garching, Germany\\
~\\}
 
\emailAdd{felix.berkhahn@physik.lmu.de}
\emailAdd{sophia.x.mueller@physik.uni-muenchen.de}
\emailAdd{florian.niedermann@physik.lmu.de}
\emailAdd{robert.bob.schneider@physik.uni-muenchen.de}

\abstract{
A theory of a non-relativistic, complex scalar field with derivatively coupled interaction terms is investigated. This toy model is considered as a prototype of a classicalizing theory and in particular of general relativity, for which the black hole constitutes a prominent example of a classicalon. Accordingly, the theory allows for a non-trivial solution of the stationary Gross-Pitaevskii equation corresponding to a black hole in the case of GR. Quantum fluctuations on this classical background are investigated within the Bogoliubov approximation. It turns out that the perturbative approach is invalidated by a high occupation of the Bogoliubov modes. Recently, it was proposed that a black hole is a Bose-Einstein condensate of gravitons that dynamically ensures to stay at the verge of a quantum phase transition. Our result is understood as an indication for that claim. Furthermore, it motivates a non-linear numerical analysis of the model. 
}

\begin{document}

\maketitle

\section{Introduction}

Recently, Dvali and Gomez proposed a microscopic picture of black holes \cite{Dvali:2011aa, Dvali:2012gb, Dvali:2012rt}. According to them, black holes can be understood as Bose-Einstein condensates of gravitons. In this picture, the Schwarzschild geometry would effectively emerge from the interaction of a test particle with the condensate of gravitons. In \cite{Dvali:2012en,Dvali:2012wq} this picture was further elaborated and the authors concluded that the black hole is at the point of quantum phase transition.

Within the Schwarzschild radius, the graviton theory is strongly coupled. This necessitates to sum up a large number of equally important terms in the perturbation series. This fact and the relativistic nature of the graviton theory makes it hard to obtain any quantitive predictions along the lines of \cite{Dvali:2011aa, Dvali:2012gb, Dvali:2012rt,Dvali:2012en,Dvali:2012wq} within the theory of general relativity. Therefore, in this paper we propose a non-relativistic, derivatively coupled toy model that allows to quantitatively compute properties expected for black holes according to  \cite{Dvali:2011aa, Dvali:2012gb, Dvali:2012rt,Dvali:2012en,Dvali:2012wq}. Our model is constructed such that it contains a ground state corresponding to the black hole of general relativity, which is nothing else but a non-relativistic classicalon state. For a description of the concept of classicalization in the case of gravity see \cite{Dvali:2010bf,Dvali:2010ue}  and for its generalisation to other derivatively coupled theories compare to \cite{Dvali:2010jz, Dvali:2011th, Dvali:2011nj,Dvali:2012zc}.  

We perform a quantum perturbation theory around a highly occupied classical state (so called 'Bogoliubov approximation') which is supposed to make up the classicalon. Our results indicate that the perturbative approach is not applicable, which is exactly what we expect to see if the system indeed manages to stay at the point of quantum phase transition. Therefore, we see indications for the claims of \cite{Dvali:2012en,Dvali:2012wq}, even though only a subsequent numerical and non-linear analysis will clearly decide about the status of our model.
 
Our paper is organized as follows: Section \ref{sec_BH} summarizes the main ideas of \cite{Dvali:2011aa}. Section~\ref{sec_model} contains our model and results. Future prospects of our theory are discussed in Section~\ref{sec_prospects}.

\section{Black Holes as Graviton Condensates} \label{sec_BH}

\subsection{Quantum Portrait}
The starting point in the approach of \cite{Dvali:2011aa, Dvali:2012gb, Dvali:2012rt,Dvali:2012en} is the observation that the graviton interaction strength $\alpha_{gr}$ is momentum dependent due to the derivatively coupled nature of interaction terms of the metric fluctuation field with itself:
\begin{equation} \label{eq_alpha_scaling}
\alpha_{gr} = h G_N \lambda^{-2}\;,
\end{equation}
where $G_N$ is Newtons constant and $\lambda$ is the typical graviton wavelength involved in a given scattering process. For the case of black holes, the characteristic wavelength is set by the Schwarzschild radius $r_g = 2\, G_N M \sim \lambda$, where $M$ is the mass of the black hole. Accordingly, each graviton contributes an energy $ \sim h / (2 \,G_N M) $. The total number $N$ of gravitons constituting a black hole is thus
\begin{equation} \label{eq_N_BH}
N = \frac{ 2 G_N M^2}{ h} \sim \frac{\lambda^2}{L_P^2}\;,
\end{equation}
where we have introduced the planck length $L_P= \sqrt{ h G_N }$. Equation (\ref{eq_N_BH}) is also true for the number of gravitons contained in the gravitational field of other objects such as planets since it can be obtained from summing up the Fourier modes of any Newtonian gravitational field $\phi = - r_g/r$. Inserting (\ref{eq_N_BH}) in (\ref{eq_alpha_scaling}) yields the dependence of the coupling with $N$
\begin{equation}\label{alpha1}
\alpha_{gr} = \frac{1}{N}\;.
\end{equation}
The occupation number $N$ can be understood as the parameter measuring the classicality of a given object composed out of gravitons, in this case black holes. Intrinsic quantum processes such as the decay into a two particle state are exponentially suppressed $\langle {\rm Out} | \exp{(-S)} |{\rm In} \rangle  \sim \exp{(-N)}$. Additionally, the number of gravitons produced in the gravitational field of any elementary particle is negligibly small, for example for an electron we get $N = 2 G_n m_e^2 / h \approx 10^{-44}$. This shows why elementary particles cannot be considered as a classical gravitating object (even though they contribute a standard Newton law at large distances), and in particular it becomes clear why a single elementary particle does not collapse into a black hole.

Let us contrast black holes with the gravitational field of other objects such as planets. Assuming that the characteristic wavelength of the gravitons is in any case given by the characteristic size $R$ of the object, we obtain as the gravitational part of the energy
\begin{equation}
E_{grav} \sim \frac{N h}{R} \sim M \frac{ r_g}{ R}\;.
\end{equation}
This shows that for objects not being a black hole (i.e., for $R > r_g$) a substantial part of the energy is carried by other constituents than gravitons. This is why the gravitational field of other objects than black holes cannot exist without an external source, for example a planet. However, once the extension of the gravitational object reaches $R=r_g$, the whole energy $M$ of our object is stored in the gravitational field, so that an external source is not required to balance the energy budget. It is exactly at this point where the interaction of an individual graviton with the collective potential generated by the other gravitons becomes significant. This can most easily be seen by appreciating that the classical perturbation series in the metric fluctuation field $h$ about a Minkowski background breaks down at the horizon~$r_g$. However, the interaction of two individual gravitons is still small as long as we consider regions $r>L_P$. Given that the dominant interaction is gravity itself, the authors of~\cite{Dvali:2011aa} concluded that black holes are \textit{self-sustained} bound states of gravitons. Moreover, black holes are \textit{maximally packed} in the sense that the only characteristic of a black hole in the semi-classical limit is the number of gravitons $N$ composing it, and any further increase of this number results inevitably in an increase of the size and mass of the black hole. This becomes clear since by default the extension of the black hole is no free parameter but given by $r_g$, and accordingly all physical black hole quantities (mass, size, entropy, etc.) can be quantified by $N$. This is nothing else but the famous no-hair theorem translated in the language of gravitons. An important consequence of this picture is that black holes always balance on the verge of self-sustainability, since the kinetic energy $h/r_g$ of a single graviton is just as large as the collective binding potential $-\alpha_{gr} N h / r_g$ produced by the remaining $N-1$ gravitons. Thus, if you give a graviton just a slight amount of extra energy, its kinetic energy will be above the escape energy of the bound state. In \cite{Dvali:2011aa} it was therefore concluded that black holes are \textit{leaky condensates}.

The above reasoning strictly applies only in the (semi-)classical limit $N \rightarrow \infty $. This is important, because we might wonder how a quantum effect like Hawking radiation can be understood in our picture of highly occupied graviton states, since usually we expect quantum effects to be exponentially suppressed. Actually, to explain this, the authors in \cite{Dvali:2012en} conjectured that the black hole is at a point of \textit{quantum phase transition}. Thus, quantum excitations are always significant and cannot be ignored. In particular, given that black holes are leaky condensates, every quantum excitation will lead to the escape of the corresponding graviton. These escaped particles are interpreted as the Hawking radiation of the black hole. 

Moreover, due to the quantum phase transition, the leading corrections to the above (semi-)classical ($N \rightarrow \infty$) picture are not exponentially but only $1/N$ suppressed. This makes it possible for any finite $N$ to retrieve information from the black hole (for instance the Hawking spectrum contains $1/N$ corrections, making it for example possible to read out the amount of Baryons originally stored in the black hole). The famous information paradox is thus just a relict of working in the strict (semi-)classical $N \rightarrow \infty $ approach in which the hair of the black hole is negligible compared to the $N$ graviton state.

In the next section we discuss the well known physics of quantum phase transition for the example of a non-relativistic condensed matter system. Assuming that black holes behave similar to this model, we will qualitatively discuss the implications for black hole physics, as it was done in \cite{Dvali:2012en}.

\subsection{On the Verge of Quantum Phase Transition} \label{sec_quantumphase}

The discussion of this section closely follows \cite{Kanamoto}, where the properties of a quantum phase transition are studied. 
We want to describe a system of $N$ bosons of mass $m$ with an attractive interaction in one dimension of size $V$ at zero temperature. The second quantized field~$\hat\Psi(x,t)$ in the Heisenberg representation is measuring the particle density at position $x$. The corresponding Hamiltonian reads
\begin{align} \label{eq_standard_BEC}
\hat H =  \frac{\hbar^2}{2m} \int_0^V {\rm d}x (\partial_x \hat\Psi)^{\dagger}  (\partial_x \hat\Psi) - \frac{U}{2} \int_0^V {\rm d}x\,  \hat\Psi^{\dagger}\hat\Psi^{\dagger}  \hat\Psi \hat\Psi \;,
\end{align}
where $U$ is a positive parameter of dimension $[\rm energy]\times[length]$ controlling the interaction strength. 
The dynamics of $\hat\Psi(x,t)$ are given by the Heisenberg equation
\begin{align} \label{eq_heisenberg}
i \hbar \frac{\partial}{\partial t} \hat \Psi 	&=  \left[ \hat\Psi,\hat H \right]\\
				&= \left( -\frac{\hbar^2}{2m}\partial_x^2 - U (\hat\Psi^{\dagger}\hat\Psi)  \right)\hat\Psi
\end{align}
where the equal time commutation relations
\begin{align}
\left[\hat \Psi(x,t),\hat\Psi^{\dagger}(x',t) \right]&=\delta(x-x') &\left[\hat \Psi(x,t),\hat\Psi(x',t) \right]&=0
\end{align}
have been used. Applying the mean-field approximation amounts to replacing the operator~$\hat\Psi(x,t)$ by a classical field $\Psi_0(x,t)$. This replacement is justified when the quantum ground state is highly occupied. In this case the non-commutativity of the field operator is a negligible effect. Since we are looking for stationary solutions, the time dependence is separated in the usual way
\begin{align}\label{sep_t}
\Psi_0(x,t)=\Psi_0(x) \exp{\left(-\frac{i\mu t}{\hbar}\right)}\;,
\end{align}
where $\mu$ is the chemical potential.
Inserting this ansatz in \eqref{eq_heisenberg}, yields the stationary Gross-Pitaevskii equation.
A trivial solution that fulfils the periodic boundary conditions $\Psi_0(0)=\Psi_0(V)$ is given by
\begin{align}\label{BEC_Sol}
\Psi^{(\rm BE)}_0(x)=\sqrt{\frac{N}{V}}=const.
\end{align} 
This solution corresponds to the homogenous Bose-Einstein condensate. However, this solution is the minimal energy configuration only for $U<U_c$. The critical value has been be derived in \cite{Kanamoto} to be: $U_c=\hbar^2 \pi^2/(mVN)$. For $U>U_c$ the ground state is given by an inhomogenous solution $\Psi^{(\rm sol)}_0(x)$ describing a soliton. By increasing the parameter $U$, i.e. the interaction strength, the ground state of the system undergoes a phase transition from the Bose-Einstein phase to the soliton phase once the critical point $U_c$ is reached. As the authors in~\cite{Kanamoto} have shown, this point of phase transition is characterized by a cusp in the chemical potential $\mu(U)$, the kinetic energy $\epsilon_{kin}(U)$ and the interaction energy $\epsilon_{int}(U)$ per particle as functions of $U$.

The main result of \cite{Kanamoto} was to show that at the point of phase transition quantum corrections to $\Psi_0$ become important and a purely classical description is no longer possible, therefrom the name 'quantum phase transition'. A suitable way to investigate this effect is provided by the Bogoliubov approximation in which the classical field $\Psi_0$ is furnished with small quantum corrections $\delta\hat\Psi$. A proper quantum mechanical treatment, of which the details are given in the next section, allows to derive the famous energy spectrum of the Bogoliubov excitations
\begin{align}
\epsilon(k) \label{b_spectrum}
&=\left(\left(\frac{\hbar^2 \delta k^2}{2m}\right)^2-\frac{\hbar^2 U N} {m V}\delta k^2\right)^{1/2}\\\notag
&=\left(\left(\frac{\pi\hbar^2 }{m V}\right)^2\delta k^2\left[\left(\frac{V}{2\pi}\right)^2 \delta k^2-\frac{U} {U_c}\right]\right)^{1/2} \;.
\end{align}
Due to the periodic boundary conditions, the momentum $\delta  k$ of the Bogoliubov modes is quantized in steps of $2\pi/V$. From \eqref{b_spectrum} it is clear that once the interaction strength approaches the value $U_c$, the energy of the first Bogoliubov mode ($\delta k=2\pi/V$) vanishes. Consequently, the excitation of the first mode becomes energetically favourable and the condensate is depleting very efficiently. This is the characteristic property of a quantum phase transition. This picture is further substantiated by calculating the occupation number of excited Bogoliubov modes
\begin{align}\label{occupationBog}
n(\delta k)=\frac{\hbar^2\delta k^2/2m - U N / V}{2 \epsilon(\delta k)}-\frac{1}{2}\;,
\end{align} 
which shows that the vanishing of $\epsilon(\delta k)$ is accompanied by an extensive occupation of the corresponding quantum states. This means that the Bogoliubov approximation is no longer applicable and quantum corrections are significant. For values $U>U_c$ the energy becomes imaginary, which signals the formation of a new ground state that is given by the soliton solution~$\Psi_{0}^{\rm (sol)}(x)$, compare to the discussion in \cite{Kanamoto}. Moreover, the work of \cite{Qian,Flassig:2012re} shows that the system becomes drastically quantum entangled at the critical point, which is yet another characterization of quantum phase transition.

By making the $N$ dependence of $U_c$ explicit and introducing the new dimensionless coupling parameter $\alpha=U m V/(\hbar^2\pi^2)$, the condition for the breakdown of the Bogoliubov approximation becomes
\begin{align}\label{alpha2}
\alpha=\frac{1}{N}\;.
\end{align} 
This is exactly the condition for self-sustainability in the case of a black hole~\eqref{alpha1}. These considerations closely follow \cite{Dvali:2012en}, where the authors wanted to illustrate the relation between black hole physics and Bose-Einstein condensation at the critical point. Of course, in this toy model the relation \eqref{alpha2} is not generically realized, but has to be imposed by adjusting the model parameters by hand. (For a given value of $N$, the interaction strength $U$ has to be chosen appropriately.) In the case of GR the left hand side of equation \eqref{alpha2} is $k$-dependant which in principal could allow for a generic cancelation between the two terms in the squared bracket in the last line of \eqref{b_spectrum}. This cancelation is assumed to take place up to $1/ N$--corrections.

The aim of our work is to present a non-relativistic scalar model that is in principle able to account for this cancelation and thus generically stays at the point of quantum phase transition independent of the chosen parameters. It is not possible to derive this result within the Bogoliubov approximation since a high occupation of quantum states is the defining property of a quantum phase transition. However, the breakdown of the perturbative approach is a necessary condition and therefore provides an indication for it.

\section{Microscopic Picture of Non-Relativistic Classicalons}
\label{sec_model}

\subsection{The Model}
Non-relativistic classicalizing theories have the advantage of being computable without a resummation of infinitely many equally important terms as it would be the case for example in GR. In the following, we will consider a special non-relativistic, classicalizing theory that was constructed to mimic general relativity.
As in \cite{Kanamoto}, we choose to confine our theory in a 1-dimensional box of size $V$. 
To be concrete, we consider the following Hamiltonian for the second quantized field $\hat \Psi(x)$ measuring the particle density at position $x$:
\begin{multline} \label{eq_model}
\hat H =  \frac{\hbar^2}{2m} \int_0^V \!\!\!{\rm d}x :\!(\partial_x \hat\Psi)^{\dagger}  (\partial_x \hat\Psi)\!: + \lambda \int_0^V \!\!\!{\rm d}x :\!\left( (\partial_x \hat\Psi)^{\dagger}  (\partial_x \hat\Psi) \right)^2\!: + \\
\kappa \int_0^V \!\!\!{\rm d}x :\!\left( (\partial_x \hat\Psi)^{\dagger}  (\partial_x \hat\Psi) \right)^3\!:\;,
\end{multline}
where $:\;:$ denotes the normal ordering. We are looking for homogenous solutions of the Heisenberg equation
\begin{equation} \label{eq_heisenberg2}
i\hbar \frac{\partial}{\partial t} \hat \Psi = \left[\hat \Psi, \hat H \right]\;,
\end{equation}
in which the field operator is again replaced by a classical field $\Psi_0(x)$. (The subscript $0$ will be suppressed throughout the rest of this work.) We try to generalize the known homogenous BEC solution \eqref{BEC_Sol}. We can separate the time dependence as in \eqref{sep_t}. Since $\Psi(x)$ is a complex field,~(\ref{eq_heisenberg2}) has in general the following class of solutions
\begin{equation} \label{eq_BEC_solution}
\Psi_{k}(x) = \sqrt{\frac{N}{V}} \exp{\left(i k x\right)}\;,
\end{equation}
where the momentum $k$ is quantized in steps of $2\pi/V$ by implementing periodic boundary conditions. The number of particles is denoted by $N$.
Inserting (\ref{eq_BEC_solution}) in the Hamiltonian (\ref{eq_model}) results in the polynomial 
\begin{equation} \label{eq_energy}
\frac{H^{(0)}}{V} =\frac{\hbar^2}{2m} z + \lambda z^2 + \kappa z^3
\end{equation}
where $z=\frac{N}{V} k^2$. 

However, not every solution (\ref{eq_BEC_solution}) is a local minimum of the energy (\ref{eq_energy}). For sure, one minimum is given by $k=0$ (since the kinetic energy contributes positively), which would exactly correspond to the Minkowski vacuum in the case of general relativity given that this is the global energetic minimum of the theory (\ref{eq_model}). Moreover, by appropriately choosing the coefficients $\lambda$ and $\kappa$, we can construct a second minimum of (\ref{eq_energy}) at $z_0=N k_0^2 /V$ with positive energy, denoted with $\Psi_{k_0}$, where $k_0>0$. It is easy to show that the corresponding solution not only minimizes (\ref{eq_energy}) (that is, minimizing the energy within the sub-class of homogenous solutions (\ref{eq_BEC_solution})) but is also given as a minimum in complete field space (that is, it is a minimum for general fluctuations $\Psi = \Psi_{k_0} + \delta \Psi$). It is this solution that will turn into the classicalon which corresponds to the black hole solution of general relativity. Furthermore, it should be noted that the chemical potential is zero due to the relation  $\mu \propto \partial H^{(0)}/\partial z|_{z_0}$.

\subsection{Bogoliubov Theory}
We will study the leading quantum perturbations $\delta\hat \Psi(x)$ about the classical condensate~$\Psi_{k_0}(x)$. To this end, we  write
\begin{eqnarray}
\hat \Psi(x) &=& \frac{1}{\sqrt{V}} \sum_k \hat a(k) e^{ikx} = \frac{1}{\sqrt{V}} \hat a(k_0) e^{ik_0x} +    \frac{1}{\sqrt{V}} \sum_{k \neq k_0} \hat a(k) e^{ikx} \label{eq_fourier}\;, 
\end{eqnarray}
where $\hat a(k)$ is the annihilation operator of the momentum mode $k$. The Bogoliubov approximation consists in treating the first term in \eqref{eq_fourier} classically due to the large occupation of the state with momentum $k_0$. The second term presents a small quantum correction. On account of this, the replacement 
\begin{align}
\hat a (k_0) \to \sqrt{N_0}
\end{align}
is introduced, which allows to identify $\Psi_{k_0}(x)$ with the first term in \eqref{eq_fourier}. The second term is simply the Fourier representation of the quantum perturbation $\delta\hat \Psi(x)$ . We want to calculate the perturbation series up to second order in $\delta\hat\Psi(x)$ or $ \hat{a}(k \neq k_0)$. Note that once we allow for an occupation of the momentum states with $k \neq k_0 $, we have to distinguish between~$N_0$, the number of particles in the ground state, and $N$, the total number of particles. Since we want to express everything in terms of $N$, the normalisation condition 
\begin{align}\label{normalization}
\hat a^{\dagger}(k_0) \hat a(k_0)=N-\sum_{k \neq k_0}\hat a^\dagger(k)\hat a(k)
\end{align}
has to be employed. This means that the zeroth order $H^{(0)}$ terms contribute to the second order $H^{(2)}$ when we express $N_0$ in terms of $N$. Inserting (\ref{eq_fourier}) and (\ref{normalization}) into the Hamiltonian (\ref{eq_model}), results in the following quadratic order expression:
\begin{eqnarray} \label{eq_hamiltonian_bogo}
H^{(2)} =  \sum_{\delta k \neq 0} \left[ \epsilon_0^{(1)} \hat a^\dagger \hat a + \epsilon_0^{(2)} \hat b^\dagger \hat b + \epsilon_1 (\hat a^\dagger \hat b^\dagger + \hat b \hat a)\right]\;,
\end{eqnarray}
where the decomposition $k=k_0+\delta k$ has been used and the (re-)definitions
\begin{align}
\hat a(\delta k) &\equiv \hat a(k_0 + \delta k)\;, \\
\hat b(\delta k) &\equiv \hat a(k_0 - \delta k) \;, \label{intro_b}
\end{align}
as well as
\begin{align}
\epsilon_0^{(1)} &= (k_0 + \delta k)^2 P_0+\Lambda_0\;, \label{eps0p}\\
\epsilon_0^{(2)} &= (k_0 - \delta k)^2 P_0 +\Lambda_0\;,\label{eps0m}\\
\epsilon_1 &= (k_0^2 - \delta k^2) P_1\label{eps1}\;,
\end{align}
apply.
Here, the polynomials $P_0$, $P_1$ and $\Lambda_0$ are functions of the combination $z_0$ and the coefficients $m$, $\lambda$ and $\kappa$:
\begin{eqnarray}
P_0 &=& \frac{\hbar^2}{4m} + 2 \lambda z_0 + \frac{9}{2} \kappa z_0^2 \label{eq_P0} \\
\Lambda_0&=&-k_0^2\left(\frac{\hbar }{4m} + \lambda  z_0+ \frac{3}{2} \kappa  z_0^2 \right) \\
P_1 &=& \lambda z_0 + 3 \kappa z_0^2 \label{eq_P1}
\end{eqnarray}
Note that when using the minimal energy condition $\partial H^{(0)} / \partial z|_{z_0}= 0$, see equation (\ref{eq_energy}), we obtain $P_0 = P_1$ and $\Lambda_0=0$ due to the relations $2 V(P_0 - P_1) = \partial H^{(0)}/\partial z|_{z_0}$  and $2 V\Lambda_0 / k_0^2=  -\partial H^{(0)}/\partial z|_{z_0}$, respectively. Furthermore, it can be checked that $P_0>0$ if $z_0$ corresponds to the minimum of \eqref{eq_energy} because $2V P_1 = z_0\;\partial^2 H^{(0)}/\partial z^2|_{z_0}$. 
The Hamiltonian (\ref{eq_hamiltonian_bogo}) is almost of the Bogoliubov form and can be diagonolised by means of the transformation
\begin{align}\label{bog_trafo}
\hat \alpha &= u \hat a + v \hat b^\dagger &\text{and}&&\hat \beta = u \hat b + v \hat a^\dagger,
\end{align}
where $u,v \in  \mathbb{R}$.
Setting the off-diagonal terms to zero and requiring standard commutation relations for $\hat \alpha$ and $\hat \beta$ implies 
\begin{align}\label{offDiag}
\epsilon_1\left(u^2+v^2\right)-2 u\, v \frac{\epsilon_0^{(1)}+\epsilon_0^{(2)}}{2}=0\;,
\end{align}
as well as
\begin{align}
u^2-v^2=1\;.
\end{align}
These two equations are solved by
\begin{align}\label{uvBog}
u = \pm \frac{1}{\sqrt{2}} \left(\frac{1}{2} \frac{\epsilon_0^{(1)} + \epsilon_0^{(2)}}{\epsilon} + 1 \right)^{1/2} \!\!,\;\;
v = \pm\frac{1}{\sqrt{2}} \left(\frac{1}{2} \frac{\epsilon_0^{(1)} + \epsilon_0^{(2)}}{\epsilon} - 1 \right)^{1/2}\!\!,
\end{align}
where 
\begin{equation}\label{eq_esp}
\epsilon=\sqrt{\frac{1}{4}\left(\epsilon_0^{(1)} + \epsilon_0^{(2)}\right)^2-\epsilon_1^2}\;.
\end{equation}
Note that $\epsilon_0^{(1)}$ and $\epsilon_0^{(2)}$ are strictly positive, whereas the sign of $\epsilon_1$ depends on the value of $\delta k$. Thus in order to fulfill \eqref{offDiag}, we have to choose $u$ and $v$ in \eqref{uvBog} both positive when $\delta k < k_0$ and one of both has to be chosen negative when $\delta k > k_0$. In both cases the diagonalized version of \eqref{eq_hamiltonian_bogo} reads
\begin{align} \label{eq_H2_diagonal0}
H^{(2)} = \sum_{\delta k \neq 0} \left[ \left(\epsilon + \frac{1}{2} (\epsilon_0^{(1)} - \epsilon_0^{(2)}) \right) \hat \alpha^\dagger \hat \alpha+ 
\left(\epsilon - \frac{1}{2} (\epsilon_0^{(1)} - \epsilon_0^{(2)}) \right) \hat \beta^\dagger \hat \beta  + \epsilon - \frac{1}{2} (\epsilon_0^{(1)} + \epsilon_0^{(2)}) \right]\;.
\end{align}
Using the definitions \eqref{eps0p}, \eqref{eps0m} and \eqref{eps1}, we find $\epsilon=2P_0k_0|\delta k|$  and  $(\epsilon_0^{(1)} - \epsilon_0^{(2)})/2=2P_0k_0 \delta k$. Note that $\epsilon$ is strictly positive. By employing the relation $\hat\alpha(\delta k) = \hat\beta(-\delta k)$ we find
\begin{align} \label{eq_H2_diagonal}
H^{(2)} = \sum_{\delta k \neq 0} \left[ 2\left(\epsilon + \frac{1}{2} (\epsilon_0^{(1)} - \epsilon_0^{(2)}) \right) \hat \alpha^\dagger \hat \alpha+ \epsilon - \frac{1}{2} (\epsilon_0^{(1)} + \epsilon_0^{(2)}) \right]\;.
\end{align}
Accordingly, the vacuum $|0\rangle$ of the Fock space is defined as 
\begin{eqnarray}
\hat \alpha |0\rangle &=& 0 \;.
\end{eqnarray}
It follows from the Hamiltonian (\ref{eq_H2_diagonal}) that the combination 
\begin{align}\label{energy_qparticles}
e(\delta k) \equiv 2\left(\epsilon + \frac{1}{2} (\epsilon_0^{(1)} - \epsilon_0^{(2)})\right)
\end{align}
is the energy of the quasi particles created by $\hat\alpha^\dagger(\delta k)$ with momentum $k_0 + \delta k$. Since the vacuum of our theory is defined with respect to $\hat \alpha$, it contains a non-vanishing amount of excited real particles associated with $\hat a$ (and $\hat b$ equivalently). This effect goes under the name quantum depletion and occurs physically due to the interactions amongst the particles which necessarily pushes some of them to excited states. Their precise number is given by
\begin{eqnarray}
\langle 0| \hat a^\dagger(\delta k) \hat a(\delta k) |0\rangle = v^2(\delta k)\;.
\end{eqnarray}
This allows to rewrite the energy of the quasi particles associated with $\hat \alpha$ as 
\begin{equation} \label{eq_gap}
e(\delta k)=\begin{cases} 
8 P_0 k_0\, \delta k\; & \text{for } \delta k > 0\\
0& \text{for } \delta k \leq 0 \\
\end{cases}
\end{equation}
and the number of depleted real particles with momentum $k_0+\delta k$ as
\begin{equation}  \label{eq_depleted}
v^2(\delta k)=\frac{1}{2} \left( \frac{k_0^2 + \delta k^2}{2k_0 |\delta k|} - 1 \right)\;.
\end{equation}
The above results can easily be generalized to a derivatively coupled theory with an arbitrary number of higher order terms
\begin{equation}\label{Hamilton_gen}
H =  \sum_{r=1}^{r_{\rm max}} c_{r}\int_0^V\! \! {\rm d}x\,  :\!(\partial_x \Psi^{\dagger} \partial_x \Psi)^{r}\!:\;.
\end{equation}
Note that the coefficients $c_{r}$ have dimension $[\rm energy][length]^{3r-1}$. The standard kinetic term corresponds to $r=1$ for which the coefficient is $c_{1}=\hbar^2/(2m)$. The energy of the quasi particles and the number of depleted particles are given by \eqref{eq_gap} and \eqref{eq_depleted} where $P_0$ now is given by the generalized expression 
\begin{align}
P_0 &=  \sum_{r=1}^{r_{\rm max}}c_{r} \,\frac{r^2}{2} \left(\frac{N}{V}\right)^{r-1} \left(k_0^2\right)^{r-1} \label{P_0_gen}\;,
\end{align}
and $k_0$ is determined as a minimum of the generalized version of \eqref{eq_energy}
\begin{equation} \label{eq_energy_gen}
\frac{H^{(0)}}{V} = \sum_{r=1}^{r_{\rm max}}c_{r}\, \left( k^2\right)^{r}\left(\frac{N}{V}\right)^r\;.
\end{equation}
The coefficients $c_{r}$ have again to be chosen such that there is a non trivial minimum.
\subsection{Discussion} \label{consequences}

Our results incorporate the vanishing of the energy gap for $\delta k < 0$. This (at least partly) vanishing energy gap can be considered as an indication for the occurrence of a quantum phase transition, as we discussed in section \ref{sec_quantumphase}. Moreover, we see that the Bogoliubov modes become highly occupied for $\delta k \gg k_0$. This in fact signals a breakdown of the Bogoliubov theory anyways, as two succeeding terms in the quantum perturbation theory compare as 
\begin{equation} \label{eq_breakdownpert}
N_0 \left(k_0 + \delta k \right)^2 k_0^2 \delta N \sim N_0^{1/2} \left(k_0 + \delta k \right)^3 k_0 \delta N^{3/2}\;,
\end{equation}
where $\delta N$ denotes the number of excited particles in the momentum state $k_0 + \delta k$. Equation~(\ref{eq_breakdownpert}) clearly shows that the number of excited particles should at least be suppressed as $\delta N \sim  N_0 k_0^2/\delta k^2$. The result for the number of depleted particles (\ref{eq_depleted}) is, however, completely the opposite, as it is not suppressed but enhanced for large $\delta k$. Therefore, we can safely conclude that the perturbative approximation has broken down anyways. Again, this is in accordance with the expectation of being at the quantum critical point because at this point the system behaves purely quantum and cannot even approximately be described classically. Therefore, the breakdown of the Bogoliubov theory was expected, since it amounts to calculate the perturbative quantum corrections around a classical ground state. 

Note that the breakdown is also intuitive from the viewpoint of a vanishing energy gap for the quasi particles with $\delta k < 0$. Of course, neither $\hat a$ or $\hat b$ particles can directly be related with the direction of $\hat \alpha$ or $\hat \beta$ particles in phase space. But the vanishing of the energy gap should somehow be transferred into the sector of physical $\hat a$ and $\hat b$ particles. Since a vanishing energy gap means that it is indefinitely easy to excite the quasi particles, we seem to recover this behavior in the high momentum sector of $\hat a$ and $\hat b$ particles.

We can also perform the Bogoliubov approximation around the global minimum of \eqref{eq_energy} at $k=0$. Due to the derivatively coupled nature of the interaction terms, the higher order terms in~\eqref{eq_model} do not contribute, which in turn implies that the Hamiltonian~\eqref{eq_hamiltonian_bogo} is already diagonal. Therefore, there is no depletion of the vacuum which allows us to further extend the GR analogy: This state would simply correspond to the Minkowski vacuum in the case of GR. 

\section{Future Prospects}\label{sec_prospects}
Contrary to model~\eqref{eq_standard_BEC}, where the critical point is actually reached and crossed by sufficiently increasing the interaction strength $U$, in our model there is some indication that the system stays at the point of quantum phase transition and does not organize itself in a new classical ground state. However, this indication is only inferred from the observation of the breakdown of the Bogoliubov theory. To get some solid measures, we need to go beyond the Bogoliubov approximation in the next step \cite{Berkhahn:appearsoon}. This can be achieved by a full quantum mechanical treatment of the theory \eqref{eq_model}. The diagonalization of the Hamiltonian can be performed under the assumption that only the lowest~$l$ momentum eigenstates are significantly occupied (given that we are supposed to sit in a local minimum, this seems to be a good assumption). Therefore, it suffices to diagonalize the Hamiltonian within a Hilbert subspace containing only a finite number of states describing $N$ bosons occupying $l$ different momentum eigenstates. For $l$ chosen appropriately small the calculation is numerically feasible and has been performed in the case of the non-derivativly coupled model in~\cite{Kanamoto}. By means of this calculation we would be able to address quantitative questions, such as the size of the energy gap, the number and spectrum of depleted particles or the amount of quantum entanglement in the system.

The generalization of our results to a relativistic classicalon theory offers another promising prospect of future research. This necessitates to apply the ideas of the Bogoliubov approach to a relativistic theory and would be a significant step towards a more quantitative treatment of the black hole condensate in general relativity.

\section*{Acknowledgements}
The authors would like to thank Gia Dvali, Daniel Flassig, Stefan Hofmann, Michael Kopp, Florian K\"uhnel, Alexander Pritzel and Nico Wintergerst for inspiring discussions. The work of FB was supported by TRR 33 'The Dark Universe'. The work of SM was supported by a research grant of the Max Planck Society. The work of FN and RS was supported by the DFG cluster of excellence 'Origin and Structure of the Universe'.

\bibliographystyle{unsrt}

\end{document}